\documentclass[aps,groupedaddress,superscriptaddress,amsmath,amssymb,twocolumn,pre]{revtex4-2}
\usepackage[dvips]{graphicx}
\usepackage{latexsym}
\usepackage{amsthm}
\usepackage{amsfonts}
\usepackage{bm}

\usepackage[colorlinks=true,citecolor=blue,linkcolor=blue,urlcolor=blue, linktocpage=true,breaklinks=true,pagebackref=false]{hyperref}

\usepackage{textcomp} 		
\usepackage{hyperref} 		
\usepackage{ifthen} 		
\usepackage{xifthen} 		
\usepackage{color} 			
\usepackage{soul} 			
\usepackage{lineno}
\usepackage[normalem]{ulem}
\usepackage{siunitx}
\usepackage{xcolor} 
\usepackage{float}
\usepackage{enumitem}

\usepackage{physics}

\def\0{^{(0)}}
\def\1{^{(1)}}
\def\2{^{(2)}}
\def\sgn{\mathrm{sgn}}
\usepackage{braket}

\newcommand{\uproman}[1]{\uppercase\expandafter{\romannumeral#1}}

\begin{document}

\title{Zero-temperature phase-flip rate in a biased parametric oscillator}
\author{D.\,K.\,J. Boneß}
\affiliation{Department of Physics, University of Konstanz, 78464 Konstanz, Germany}
\author{W. Belzig}
\affiliation{Department of Physics, University of Konstanz, 78464 Konstanz, Germany}
\author{M.\,I. Dykman}
\affiliation{Michigan State University, East Lansing, MI 48824, USA}
%

\date{\today}
	
\begin{abstract} 

A parametrically driven  oscillator has two stable vibrational states at half the modulation frequency. The states have opposite phase and equal amplitudes. An extra drive at half the modulation frequency provides an effective bias that lifts the state symmetry. Quantum fluctuations lead to switching between the states, i.e., to phase-flip transitions. We develop a semiclassical approach that allows us to find the dependence of the switching rates on the amplitude of the bias and the parameters of the modulating field. We  find that the rate of switching from a ``shallow'' state can become anomalously small at certain parameter values, leading to an efficient localization in this state. This is a consequence of the change of the topology of the oscillator phase trajectories. The results pave the way for implementing nonreciprocal quantum Ising systems based on parametric oscillators.

\end{abstract}
\maketitle

\section{INTRODUCTION}

Recent interest in quantum parametric oscillators is largely motivated by the possibility to use their states as qubits and in quantum metrology \cite{Leghtas2013,Mirrahimi2014,Goto2016,Zhang2017b,Puri2017e,Grimm2020,Zilberberg2023,Chavez-Carlos2023,Venkatraman2024,Reglade2024,Guo2024a,Marquet2024}.  This possibility is a consequence of the symmetry of  parametrically excited vibrations, which occur at half the modulation frequency $\omega_p$. Classically, the vibrations have equal amplitudes and opposite phases. Their symmetry is seen from a simple argument: incrementing time by the modulation period $2\pi/\omega_p$ does not change the modulation, but corresponds to incrementing the phase of vibrations at frequency $\omega_p/2$ by $\pi$ and thus to changing from one vibrational state to another. Symmetric vibrational states of a quantum parametric oscillator are generalized coherent states of opposite signs. Symmetric and antisymmetric combinations of such coherent states are the states of a cat qubit \cite{Mirrahimi2014}. 

A related aspect of parametric oscillators that has been attracting much interest is the possibility to associate their two vibrational states with two spin states. This suggested using coupled parametric oscillators as  Ising machines for classical and quantum annealing \cite{Wang2013,McMahon2016,Puri2017,Goto2018, Bello2019,Yamamoto2020,Mohseni2022,Ng2022,Alvarez2024}. Other applications of parametric oscillators range from force and mass sensing \cite{Rugar1991,Karabalin2011,Eichler2018} to the studies of rare events in classical and quantum systems far from thermal equilibrium \cite{Dykman1998,Lapidus1999,Marthaler2006,*Marthaler2007,Chan2007,Chan2008a,Venkatraman2024,Frattini2024,Han2024} and phase transitions into a time-symmetry-broken (time-crystal) state \cite{Kim2006,*Heo2010,Dykman2018,Heugel2019a}, see \cite{Eichler2023} for a review. 

A consequence of the symmetry of the vibrational states is the possibility of quantum tunneling between them, as suggested in Ref.~\cite{Wielinga1993}. Such tunneling is reminiscent of tunneling between the quantum states at the minima of  a symmetric double-well potential. In the presence of dissipation, it leads to interwell switching. Switching between the symmetric vibrational states corresponds to a phase flip.

If oscillators are coupled to a thermal reservoir, phase flips may also occur via transitions over the effective barrier that separates the states in phase space. Such transitions are reminiscent of thermally activated transitions over a potential barrier. However, they can occur even for zero temperature. The effective activation energy is smaller than the doubled tunneling exponent, which determines the tunneling switching rate where the tunnel splitting is small compared to the level broadening.   

A natural way of controlling parametrically excited vibrations, which underlies many of their applications, is the breaking of their symmetry. It can be implemented by applying an extra force at frequency $\omega_p/2$. With this force, the system no longer has symmetry with respect to time translation by $2\pi/\omega_p$.  Classically, the effect of the force can be understood if one thinks that it is in phase with one of the vibrations and in counteraphase with the other. Then the amplitudes of the vibrations become different. 

In this paper we consider another, and a much stronger, effect of the force at $\omega_p/2$. It is the change of the effective barrier for activated interstate switching, and thus the exponential change of the phase-flip rate. The barrier changes have opposite signs for the two vibrational states. This is  similar to changes of the barrier heights for two wells of an initially symmetric  double-well  potential by a static bias, so that we can call the force at $\omega_p/2$ a {\it ``dynamical bias''}. 
In the classical regime, the change of the switching barrier was found before  to the lowest order in the amplitude of the biasing force \cite{Ryvkine2006a}. The bias-induced change of the stationary populations of the vibrational states with the opposite phases was observed in Ref.~\cite{Mahboob2010}.

In this paper we study the effect of the bias for $T=0$, where the difference from the classical regime, and in particular the qualitative difference between phase flips and thermally activated transitions between potential wells, are most pronounced. Here, as we show, the phase-flip rates of a dynamically biased parametric oscillator can be calculated in the explicit form for an arbitrary amplitude of the bias. This is different from the analysis of the effect of quantum fluctuations for nonzero $T$, which was limited to a comparatively weak bias \cite{Boness2024a}. 

The possibility to find the barrier height for $T=0$ is a consequence of the detailed balance relation between the rates of transitions  between the quantized states (the Floquet states) of the parametric oscillator. Such relation was found earlier in the absence of dynamical bias \cite{Marthaler2006}. As we show, it holds in the presence of the bias as well.  Arguably, it underlies the condition of the vanishing of the probability current in the stationary state in the coherent-state representation, which was used  in Refs.~\cite{Bartolo2016,Roberts2020,Roberts2021} to find and analyze the stationary distribution of a dynamically biased oscillator. We should comment that a parametric oscillator is a driven system, it is far from thermal equilibrium. Generally, there are no reasons for the detailed balance to hold, and indeed it does not hold for $T >0$ \cite{Marthaler2006,Boness2024a}. 

\begin{figure}[h]
	\includegraphics[width=0.8\linewidth]{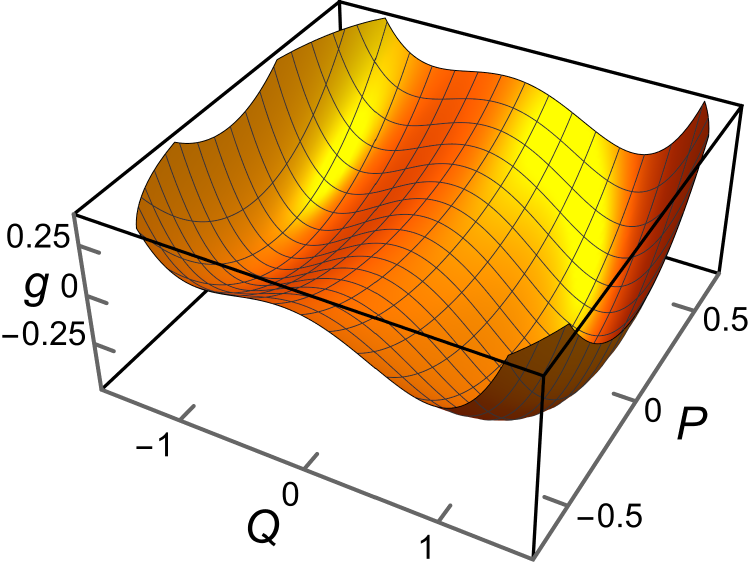}
	\caption{ Floquet (quasienergy) Hamiltonian function $g$ of a dynamically biased parametric oscillator as a function of the coordinate $Q$ and momentum $P$ in the rotating frame. In the presence of weak dissipation, the minima of the wells correspond to the stable vibrational states with opposite phases. The wells at negative and positive $Q$ are labeled in the text by $\sigma=-1$ and $\sigma=1$, respectively. The plot refers to the dimensionless value of the scaled detuning of the parametric drive $\mu=0.2$ and the scaled amplitude of the biasing field $\alpha_d = -0.1$.}\label{fig:q_energy}
\end{figure}

Our analysis is based on the semiclassical approximation. The underlying picture is that the Floquet (quasienergy) surface of the oscillator as a function of the coordinate and momentum in the rotating frame has   two wells, see Fig.~\ref{fig:q_energy}. Their minima correspond to stable vibrational states. A phase flip corresponds to switching between the wells. We assume that the oscillator has many quantized states inside the wells. Phase flip results from the random walk over these states. We find the rates of the transitions between the Floquet states, and from there evaluate the phase flip rates.    

Phase flips were recently studied experimentally in  a dynamically biased parametric oscillator based on a superconducting system \cite{deAlbornoz2024}. The observations refer to the range where the number of intrawell Floquet states is small so that tunneling plays an important role, leading to oscillations of the switching rate with the varying bias. This interesting unexpected effect is beyond the scope of the present paper, as explained below; however, our results explain the observation of the relation between the parameters where the quasienergy levels in different wells align, enabling resonant tunneling. 

In Sec.~\ref{sec:model} we introduce the Hamiltonian of the dynamically biased parametric oscillator in the rotating frame; this Hamiltonian is a double-well surface as a function of the dynamical variables. 
In Sec.~\ref{sec:topology} we discuss the classical Hamiltonian trajectories and their topological features in the complex phase space and time, including the onset of an hourglass in the phase portrait. These feature determine the rates of dissipation-induced transitions between the semiclassical intrawell Floquet eigenstates, which are calculated in Sec.~\ref{sec:activation} for zero thermal occupation number of the oscillator. We assume that the dissipation-induced level broadening is small compared to the intrawell level spacing, but is large compared to the interwell tunnel splitting. The inter-eigenstate transitions lead to a quantum walk (diffusion) over quasienergy. It forms the distribution over quasienergy, which  is studied in Sec.~\ref{sec:activation}. Ultimately the intrawell quasienergy diffusion leads to switching over the barrier that separates the wells of the Hamiltonian, and thus to phase flips. We calculate the phase flip rates in Sec.~\ref{sec:switching_rates} and reveal the possibility of localization of the oscillator in a ``shallow'' well of the Hamiltonian. Sec.~\ref{sec:conclusions} presents the summary of the results and the concluding remarks.


\section{The model}
\label{sec:model}

The dynamics of a  quantum parametric oscillator has been extensively analyzed in various contexts, see the recent work \cite{Bartolo2016,Roberts2020,Roberts2021,Kamenev2023,Thompson2023} and references therein. Of central attention has been the model of an underdamped oscillator with eigenfrequency $\omega_0$, which is modulated by a force at frequency $\omega_p$ close to $2\omega_0$, so that $|\omega_p-2\omega_0|\ll \omega_0$. The modulation is described by the term $\frac{1}{2}q_0^2F_p\cos\omega_p\bar t$ in the oscillator Hamiltonian, where $q_0$ is the oscillator coordinate and $F_p$ is the modulation amplitude; we use the notation $\bar t$ for the time in the laboratory frame. An important role for stabilizing parametrically excited vibrations is played by the Duffing (Kerr) oscillator nonlinearity, described by the term $m_0\gamma q_0^4/4$ in the Hamiltonian, where $m_0$ is the oscillator mass \cite{Landau2004a}. We consider the case where, in addition to the parametric modulation, the oscillator is driven by the force $-A_d \sin (\omega_p\bar t/2)$. With this force, the oscillator Hamiltonian has symmetry with respect to time translation by period $\pi/\omega_p$ rather than $2\pi/\omega_p$ for $A_d=0$.

The standard approach \cite{Bartolo2016,Roberts2020,Roberts2021,Kamenev2023,Thompson2023} is to describe the dynamics in the rotating frame in the rotating wave approximation (RWA). In terms of the raising and lowering operators $\hat a^\dagger$ and $\hat a$, the scaled oscillator Hamiltonian $\hat g$ reads  
\begin{align}
\label{eq:in_terms_of_ladder}
		\hat g=&  -  \lambda \mu\hat a^\dagger \hat a + \frac{\lambda}{2} (\hat a^2 + \hat a^{\dagger^2}) +\lambda^2 (\hat a^\dagger \hat a + 1/2)^2  \nonumber\\
	&- i (\lambda/2)^{1/2} \alpha_d \left(\hat a - a^\dagger \right)\,, \quad \lambda = 3\gamma \hbar/F_p \omega_p\,.
\end{align}
The parameter $\lambda$ is the  Planck constant scaled by the oscillator nonlinearity $\gamma$ and the amplitude and frequency of the modulating field, whereas $\mu$ is the scaled detuning of half the modulating field frequency from the oscillator eigenfrequency, 
\begin{align}
\label{eq:mu}
\mu = 2m_0\omega_p \delta \omega /F_p, \qquad \delta \omega = \omega_p/2 - \omega_0.
\end{align}
The scaled amplitude   $\alpha_d = A_d \sqrt{6\gamma m_0/F_p^3}$ of the force at frequency $\omega_p/2$ determines the dynamical bias. The RWA applies as long as the detuning $|\delta\omega|$ and the force parameters $F_p$ and $A_d$ are comparatively small, so that the amplitude of the parametrically excited vibrations is not large, which allows keeping only terms $\propto (a^\dagger a)^2$ in the nonlinear part of the Hamiltonian.

It is convenient to rewrite the Hamiltonian $\hat g$ in terms of the scaled coordinate and momentum of the oscillator in the rotating frame 
 \[Q=i(\lambda/2)^{1/2}(\hat a - \hat a^\dagger), \quad P=(\lambda/2)^{1/2}(\hat a + \hat a^\dagger),\]
with
\begin{align}
\label{eq:full_g}
&\hat g= \frac{1}{4} (Q^2+P^2-\mu)^2 + \frac{1}{2} (P^2 - Q^2) -\frac{\mu^2}{4}-\alpha_d Q.
\end{align}
Here and below we omit hats over $Q$ and $P$. These are operators,  $P=-i\lambda\partial_Q$, but in the classical picture they become dynamical variables. In making a transformation from Eq.~(\ref{eq:in_terms_of_ladder})  to Eq.~(\ref{eq:full_g}) we omitted a constant $\lambda\mu/2$. We note that $\mu$ should be renormalized, 
\[\mu \to \mu + 2\lambda,\]
if we define $\omega_0$ in such a way that the spacing between the ground and first excited state of the undriven oscillator is $\hbar\omega_0$. 


\subsection{Floquet states}

The Hamiltonian $\hat g$ has two parameters, $\mu$ and $\alpha_d$. It describes the oscillator dynamics in slow dimensionless time $\tau = t F_p/2m\omega_p$. The Hamiltonian function $g(Q,P)$ given by Eq.~(\ref{eq:full_g})  is shown in Fig.~\ref{fig:q_energy}.  We will be interested in the range of $\mu$ and $\alpha_d$ where this function has a two-well shape with a saddle-point at $Q=Q_s, P_s=0$, 
\[-1 + 3 (|\alpha_d|/2)^{2/3}< \mu < Q_s^2 + 1 ,\]
with $Q_s$ being the middle real root of the cubic equation $\partial_Q g(Q,0)=0$.

We label the wells at negative and positive $Q$ by  $\sigma=- 1$ and $\sigma=1$, respectively, and sometimes refer to them as the left and right wells. For $\alpha_d>0$ the deeper minimum of $g(Q,P)$, with $g=g_\mathrm{min}^\mathrm{deep}$, is located at $Q>0$ (the $\sigma=1$-well), whereas the  shallower one, with $g=g_\mathrm{min}^\mathrm{shallow}$, is located at $Q<0$. At the saddle point of $g(Q,P)$, which is on the boundary between the wells, $g=g_s$. In the presence of weak dissipation the minima of $g(Q,P)$ become stable vibrational states.  

For $\alpha_d=0$ the function $g(Q,P)$ is symmetric, it is even in $Q$ and $P$, and the wells have equal depth. The dynamical bias breaks the symmetry, and the depths of the wells become different. We note that $\hat g$ is not a sum of the kinetic and potential energies, moreover, it is quartic in the momentum $P$.

We will be interested in the parameter range where there are many eigenstates of the operator $\hat g$ inside the wells of $g(Q,P)$. This happens where the dimensionless Planck constant is small, $\lambda\ll 1$. 
The intrawell states are localized, and the matrix elements of $\hat g$ on these states are the approximate eigenvalues of $\hat g$. Even where the states in different wells have the same effective energies for certain values of $\mu$ and $\alpha_d$, the intrawell-state picture is adequate, as the overlap of the states in different wells is exponentially small and the tunnel splitting is normally much smaller than the decay rate.  Therefore the intrawell states are the approximate Floquet (quasienergy) eigenstates, and the eigenvalues of $\hat g$ are the scaled Floquet eigenvalues of the oscillator.

For small $\lambda$, the intrawell eigenstates of $\hat g$ can be described in the WKB approximation. The classical intrawell motion is periodic vibrations with a given $g$ described by the Hamiltonian equations
\begin{align}
\label{eq:eom}
\dot Q = \partial_P g(Q,P),\qquad \dot P = -\partial_Q g(Q,P),
\end{align}
where $\dot O \equiv dO/dt \equiv (dO/d\bar t)(2\omega_p/F_p)$.  In the WKB approximation, the eigenvalues $g_n$ of $\hat g$ for a given well are determined by the Bohr-Sommerfeld condition \cite{Marthaler2006}
\begin{align}
\label{eq:eigenvalues}
I(g_n) = \lambda \left(n+\frac{1}{2}\right), \quad I(g)=\frac{1}{2\pi}\oint P(Q|g) dQ.
\end{align}
Here $I(g)$ is the classical action calculated for the intrawell trajectory (\ref{eq:eom}) with a given $g(Q,P)=g$, and $P(Q|g)$ is the momentum on this trajectory. 

Equation (\ref{eq:eigenvalues}) applies to the states in each of the wells of $g(Q,P)$. Generally, in the presence of the bias $\propto \alpha_d$, the eigenvalues $g_n$ are different in different wells.
At the same time, unexpectedly, the frequency of the intrawell vibrations  
\[ \omega (g) = dI/dg\]
is the same in the both wells, for a given $g$. This is a consequence of the structure of the Hamiltonian function $g(Q,P)$. Not only in the absence of the bias \cite{Marthaler2006}, but also where $|\alpha_d|>0$, the corresponding Hamiltonian vibrations are described by the Jacobi elliptic functions, see Appendix~\ref{sec:append_trajectories}. Such functions have only one real period $\tau_p\1(g)=2\pi/\omega(g)$, which is therefore the same in the both wells. They also have a complex period $\tau_p\2(g)$.


\section{Topology of the Hamiltonian trajectories } 
\label{sec:topology}

Major qualitative features of the Hamiltonian dynamics can be understood without appealing to the explicit expressions for the Hamiltonian trajectories. In particular, the presence of at least two  periods of motion for a given quasienergy $g$ above the shallower minimum of $g(Q,P)$ can be seen already from Fig.~\ref{fig:q_energy}. One expects that, along with intrawell vibrations, there is periodic motion in complex time back and forth  between the wells in the classically forbidden region. This is reminiscent of the dynamics of a particle with mass $M$ and  energy $E$ in a static double-well potential $U(q)$. There, the underbarrier motion in the region $E<U(q)$ is oscillations in imaginary time in the inverted potential $-U(q)$. The period of these oscillations $\tau_p\2$ is  $i\sqrt{2M}\int_{q_1}^{q_2} dq/[U(q)-E]^{1/2}$, where $q_{1,2}$ are the turning points, $U(q_{1,2}) = E$, cf. \cite{Kagan1992}.  The trajectory in imaginary time $\bigl(q(t),p(t)\bigr) $ connects the real-time trajectories $\bigl(q(t),p(t)\bigr) $ in different potential wells.

\begin{figure}[h]
	\includegraphics[width=1\linewidth]{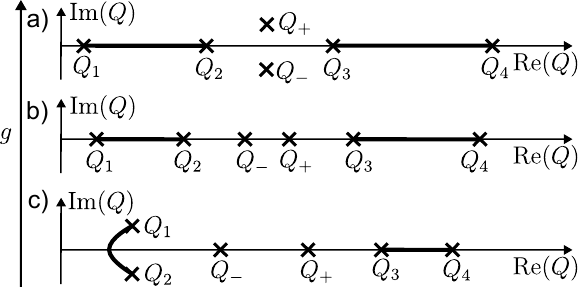}
	\caption{The classical turning points, $\dot Q =0$, in the complex plane $(\mathrm{Re}~Q, \mathrm{Im}~Q)$  for three different values of $g$. The points $Q_{1,2}$ and $Q_{3,4}$ are the turning points of motion in the left and right wells of $g(Q,P)$, respectively. They lie on the real axis for $g$ above the minimum of the shallower well $g_\mathrm{min}^\mathrm{shallow}$. The zeros $Q_\pm$ of $B(Q)$ in Eq.~(\ref{eq:turning_points}) are complex for $g>g_c$, where $g_c$ is given by Eq.~(\ref{eq:critical_g}). They merge on the real axis for $g=g_c$ and are real for $g<g_c$.  For $g<g_\mathrm{min}^\mathrm{shallow}$ the turning points $Q_{1,2}$ become complex. They are connected by a real-time trajectory in the complex-$Q$ plane.
	}
	\label{fig:sketch1}
\end{figure}

A key feature of the problem considered here is that $g(Q,P)$ is neither a sum of the kinetic and potential energies nor quadratic in $P$. As a result, the motion in the classically forbidden region is far more complicated than in the above example. A glimpse can be gained from Fig.~\ref{fig:sketch1}, which shows the turning points $\dot Q=0$ of the Hamiltonian trajectories (\ref{eq:eom})   in the complex plane $(\mathrm{Re}~Q, \mathrm{Im}~Q)$.  The points can be immediately found from the equation $\dot Q =\partial_Pg(Q,P)=0$ by noting that
\begin{align}
\label{eq:turning_points}
&\partial_Pg(Q,P)=P(Q|g)B^{1/2}(Q), \quad P^2(Q|g) = - Q^2 -1 \nonumber\\
& +\mu +  B^{1/2}(Q), \qquad B(Q) = 4(Q-Q_+)(Q-Q_-)\,,\nonumber\\
&Q_\pm = \frac{1}{2}\left\{-\alpha_d \pm [-4g+\alpha_d^2 -(1-\mu)^2]^{1/2}\right\},
\end{align}
Here, $P(Q|g)$ is the value of the momentum $P$ as a function of $Q$ for a given $g$.  We note that the equation $g(Q,P)=g$ is a quadratic equation for $P^2$; we keep the solution of this equation with $B^{1/2} >0$ on the intrawell orbits, so that $P^2(Q|g)=0$ at the boundaries of these orbits.  With Eq.~(\ref{eq:turning_points}), the equation $\partial_Pg=0$ reduces to a combination of quadratic and quartic equations for $Q$. 

\subsection{Interwell trajectories}
\label{subsec:interwell_trajectoruy}

We will first consider the range of quasienergies $g$ above the both minima of the wells of $g(Q,P)$. Here the system has two classical orbits for a given $g$. In Fig.~\ref{fig:sketch1}, these orbits lie on the real axis between $Q_{1}$ and $Q_2$, for the left well of $g(Q,P)$, and between $Q_3$ and $Q_4$, for the right well; $P(Q|g)=0$ at these points, similar to the trajectories of a particle in a double-well potential.  The region between $Q_2$ and $Q_3$ is the interwell region in Fig.~\ref{fig:q_energy}, for the corresponding $g$. However, besides the turning points $Q_{1,2,3,4}$, it follows from Eq.~(\ref{eq:turning_points})  that the system has two more turning points, $Q_+$ and $Q_-$, which may be complex.

\begin{figure*}[t]
	\includegraphics[width=1\linewidth]{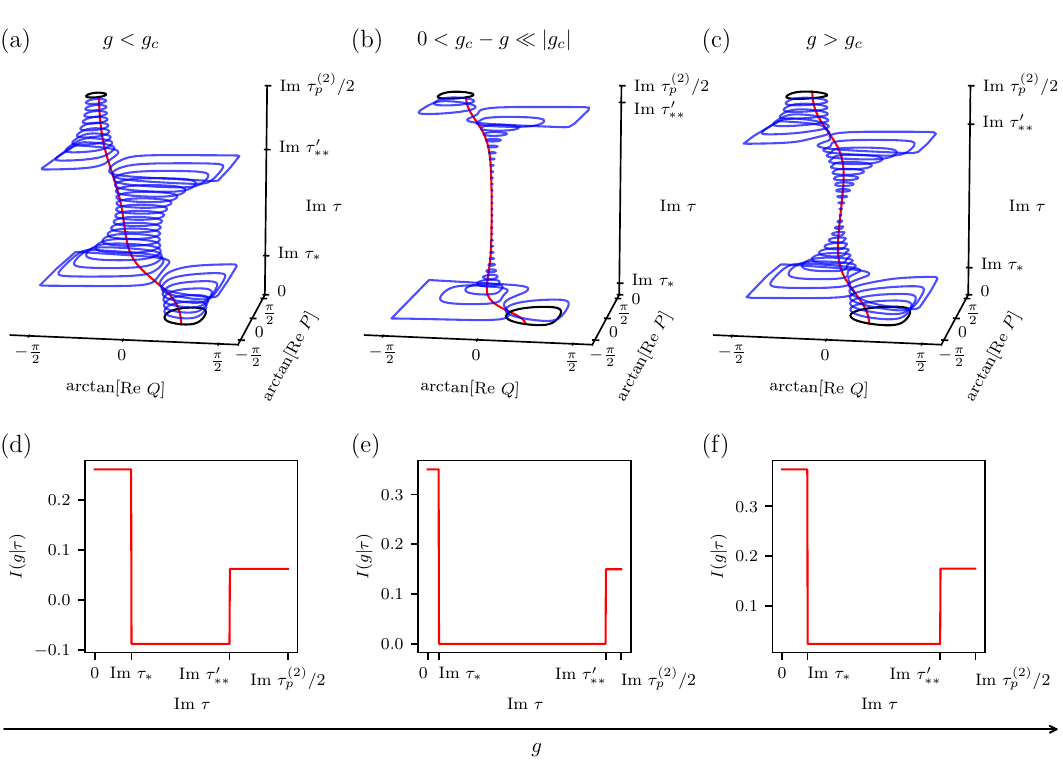}
	\caption{Phase portrait and action of the Hamiltonian system. Panels (a) - (c) show the evolution of the phase portrait with the varying complex time $\tau$ for different values of $g$. The trajectories $Q(t+\tau), P(t+\tau)$ (blue lines) are calculated from Eq.~(\ref{eq:eom}) as functions of real time $t$ for different imaginary $\tau$. We plot the arctangent of the real parts of $Q(t + \tau)$ and $P(t+\tau)$, since $Q$ and $P$ are complex for complex $\tau$ and diverge at $\tau=\tau_\ast$ and $\tau=\tau_{\ast\ast}'$. For $\tau=0$ and $\tau = \mathrm{Im}~\tau_p\2/2$, the trajectories are real and correspond to the intrawell trajectories in the wells of $g(Q,P)$ with $Q>0$ and $Q<0$ (i.e., $\sigma=1$ and $\sigma = -1$), respectively. The intrawell trajectories are shown by black lines. The imaginary-time interwell trajectories that connect the intrawell ones are shown by red lines. Panel (b) refers to $g$ close to $g_c$, where the imaginary part of the period $\tau_p\2$ is large. At $g=g_c$ the intrawell trajectories (the black lines) are disconnected. The pattern of the trajectories becomes singular, it forms a double cone. Panels (d)-(f) show the action, Eq.~(\ref{eq:action_tau}),  as a function of the imaginary time $\tau$. There are two distinct jumps at $\tau = \tau_\ast$ and $\tau = \tau_{\ast\ast}'$. For all plots $\mu=0.5$ and $\alpha_d =0.2$ such that $g_c$  is above the minima of the both wells, $g_c >g_\mathrm{min}^\mathrm{shallow} \geq g_\mathrm{min}^\mathrm{deep}$.  In the panels (a,d), (b,e) and (c,f) $g=-0.2$, $g=-0.05251$ and $g=-0.02$, respectively.}
	\label{fig:topology}
\end{figure*}

The topology of the trajectories that connect the intrawell trajectories is different in the regions  of $g$ where $Q_\pm$ are complex or real. This is illustrated in Fig.~\ref{fig:topology}. Where $Q_\pm$ are complex, Fig.~\ref{fig:sketch1}~(a) suggests that there may exist an interwell trajectory that lies on the real $Q$-axis. 
On this trajectory both $P(Q|g)$ and the travel time are purely imaginary. This is similar to a particle in a double-well potential. The travel time between $Q_2$ and $Q_3$ is half the imaginary period of interwell oscillations $\tau_p^{(2)}/2$, i.e.,
\begin{align}
\label{eq:tau_p_2}
\tau_p\2\equiv \tau_p\2(g) = 2 \int_{Q_3}^{Q_2} dQ/[P(Q|g)B^{1/2}(Q)].
\end{align}
If we count the phases on the classical intrawell trajectories off from the points where $Q=Q_2$ and $Q=Q_3$, respectively, we see that the imaginary-time trajectory connects the classical trajectories with the same phase.   

The situation is different where the turning points $Q_\pm$ lie on the real $Q$-axis, as shown in Fig.~\ref{fig:sketch1}~(b). Since $B(Q)=0$ for $Q=Q_\pm$, there is no real-$Q$ trajectory that connects the intrawell trajectories: such trajectory would be reflected back from the turning points. The connecting trajectory should go in complex-$Q$ plane around the turning points.  Near a turning point $Q_t$ ($Q_t=Q_+$ or $Q_-$)
\[ \dot Q = C(Q-Q_t)^{1/2}, \qquad |Q-Q_t|\ll 1,\]
with some (complex) constant $C=C(Q_t)$. Going around such a turning point adds a real component to the otherwise imaginary time $\int dQ/\dot Q$. This component accumulates between the turning points $Q_-$ and $Q_+$. It makes the interwell travel time $\tau_p\2/2$ complex and thus leads to a phase shift between the intrawell trajectories connected by the complex trajectory. 

The phase shift can be thought of as topological. It is determined just by the structure of $g(Q,P)$. Moreover, one may expect that this shift is equal to $\pi$. Indeed, the time to go back and forth between the points $Q_2$ and $Q_3$ is the complex period of the Hamiltonian trajectory. Since the trajectory is described by Jacobi elliptic functions, it  should bring the system to the same phase modulo $2\pi$, see also Appendix~\ref{sec:append_trajectories}; the results for the single-well regime are discussed in Appendix~\ref{sec:append_single_well}. We note that, in calculating $\tau_p\2/2$,  one should go around $Q>Q_+$ and $Q<Q_-$  on the opposite sides in the complex-$Q$-plane, so that $B^{1/2}>0$ for $Q>Q_+$ and $Q<Q_-$. 

The Hamiltonian trajectories are also described by the Jacobi elliptic functions in the range of $g$ where the function $g(Q,P)$ has only one well, i.e. for $g<g_\mathrm{min}^\mathrm{shallow}$. In general we can have $g_c\gtrless g_\mathrm{min}^\mathrm{shallow} $. The topological features for the single-well region of $g$ are discussed in Appendix~\ref{sec:append_single_well}.

\subsection{The critical point}
\label{subsec:critical_point}

The two topologically distinct  regimes refer to different values of the quasienergy $g$. They should be separated by a singularity. The singularity occurs for $g=g_c$, such that $Q_-=Q_+ = Q_c$. From Eq.~(\ref{eq:turning_points}),
\begin{align}
\label{eq:critical_g}
g_c = -\frac{1}{4}[(1-\mu)^2 - \alpha_d^2], \quad Q_c=-\alpha_d/2.
\end{align}
Since $Q_c$ is the point of merging of the two turning point, near $Q_c$ the equation of motion for $Q$ has the form $\dot Q = C'(Q-Q_c)$ with an imaginary $C'$. Then  the time of going from one intrawell trajectory to the other diverges, i.e., the trajectories become disconnected.

The change of the topology is seen in Fig.~\ref{fig:topology}.  Panels (a-c) show how the projections of the phase trajectories on the real plane $(\mathrm{Re}~Q, \mathrm{Re}~P)$ change with the varying imaginary time for three different values of $g$. The trajectories are obtained by evolving $Q(t+\tau), P(t+\tau)$ in real time $t$ over the period $2\pi/\omega(g)$, starting at different imaginary $\tau$.  The shape of the trajectories smoothly evolves with varying $\tau$ for $g$ not close to $g_c$.

As $g$ increases and approaches the critical value $g_c$, the duration of the imaginary time to connect the intrawell trajectories is increasing. The ``neck'' connecting the intrawell trajectories contracts. At $g=g_c$ the pattern of the trajectories forms a double cone with a vertex at $Q= Q_c$ and $P=0$ (we note that $\dot Q = \dot P = 0$ at this point).  For $g> g_c$ the intrawell trajectories reconnect. 

A consequence of the  nontrivial topology of the phase portrait is seen from the imaginary-time interwell trajectories shown by red lines in Fig.~\ref{fig:topology}~(a)-(c). These lines connect the intrawell trajectories  at points where $P$ has the same or opposite signs depending on whether  $g<g_c$ or $g>g_c$.  If the phases of the intrawell trajectories in the wells at $Q>0$ and $Q<0$ are counted off from the turning points $Q_3$ and $Q_2$, respectively, the change of the sign of $P$  can be shown to result from the change of the phase accumulation along the interwell trajectory equal to $\pi$. 

It is important to understand how the period $\tau_p\2$ depends on $g-g_c$ close to the divergence at $g =g_c$. As seen from Eq.~(\ref{eq:turning_points}), for small $|g-g_c|$ and small $|Q-Q_c|$ the equation of motion has the form 
\begin{align}
\label{eq:critical_Q}
&\dot Q = C'[(Q-Q_c)^2 - (g_c-g)]^{1/2},\nonumber\\
 & |Q-Q_c|\ll 1, \quad |g-g_c|\ll 1.
\end{align}
From this equation, the imaginary part of the period of motion diverges as, 
\[ \mathrm{Im}~\tau_p\2(g) \propto -\log |g-g_c|, \quad |g- g_c|\ll |g_c|.\]

In the absence of the drive at half-the modulation frequency, where $\alpha_d=0$, we have $Q_c=0$. The logarithmic divergence of the imaginary period of motion in this case was found earlier \cite{Marthaler2006} using the explicit expression for the Hamiltonian trajectories, but the topological nature of this divergence was not revealed. 
 

\subsection{Poles of the trajectories}
\label{subsec:poles}

Since the functions $Q(t)$ and  $P(t)$ are described by the Jacobi elliptic functions, they do not have branching points, but can have poles where $|Q(t)|, |P(t)|\to \infty$. The functions $Q(t), P(t)$ are double periodic, with the real period $\tau_p\1 = 2\pi/\omega(g)$ and the second period $\tau_p\2(g)$.  We will consider the positions of the poles assuming that the right well of $g(Q,P)$ is deeper, $\alpha_d >0$. We set $Q(0)=Q_3$ and $P(0)=0$ and seek the poles in the upper half of the complex-time plane within the parallelogram of periods formed by the periods $\tau_p^{(1)}$ and $\tau_p^{(2)}$. 

One pair of the poles can be easily found by integrating the equation
\[dt = dQ/\dot Q \equiv dQ/P(Q|g)B^{1/2}(Q)\]
along a trajectory that goes to $Q\to \infty$ along the real axis of $Q$. If we start at $Q=Q_3$ in Fig.~\ref{fig:sketch1}, as the system moves from $Q_3$ to $Q_4$ it acquires half the real period. For $Q>Q_4$ the momentum $P$ becomes imaginary. Taking into account  that $\dot P <0$ at $Q=Q_4$ and choosing Im~$P <0$ for $Q>Q_4$, we find that the pole is located at
\begin{align}
\label{eq:tau*}
 \tau_*(g) = \tau_p^{(1)}/2 + i\int_{Q_4}^\infty dQ/|P(Q|g)|B^{1/2}(Q). 
 \end{align}
A direct calculation shows that $\tau_* < \mathrm{Im}~\tau_p\2/2$, see Appendix~\ref{sec:append_trajectories}. We note that the calculated positions of the poles are defined modulo $\tau_p\1$. Therefore, depending on which parallelogram of periods is considered, one can replace $\tau_p\1/2$ with $-\tau_p\1/2$ in the expression for $\tau_*$. 

If we choose Im~$P> 0$ for $Q>Q_4$, the imaginary time to move to $Q\to \infty$ will be negative. The pole in the upper halfplane is found by incrementing this time by $\tau_p\2$. This gives
\[ \tau_*'(g) = \tau_p^{(1)}/2 + \tau_p^{(2)} - i\int_{Q_4}^\infty dQ/|P(Q|g)|B^{1/2}(Q). \]
It is seen from Eq.~(\ref{eq:turning_points}) that near $\tau_*(g)$ and $\tau_*'(g)$
\begin{align}
\label{eq:pole1}
&Q\approx \frac{-i}{2(t - \tau_*)}, \quad P \approx -iQ, \quad |t - \tau_*|\ll 1,\\
&Q\approx \frac{i}{2(t - \tau_*')}, \quad P \approx iQ, \quad |t - \tau_*'|\ll 1.
\end{align}

In the range of $g$ where $g(Q,P)$ has two wells, two more poles, $\tau_{**}$ and $\tau_{**}'$, can be reached by starting at $Q_3$, moving in complex time by $\tau_p^{(2)}/2$ to $Q_2$ (with Im~$P >0$ for $Q_2<Q<Q_3$), then in real time to $Q_1$, and from there moving to $Q\to -\infty$. Again, the poles can be reached by choosing Im~$P<0$ or Im~$P>0$ on the trajectories from $Q_1$ to $Q\to -\infty$. The pole $\tau_{**}$ is located at
\begin{align}
\label{eq:tau**}
	\tau_{**}(g) &= \tau_p\2/2  +\tau_p\1/2 \\
&+ \mathrm{sgn (Im}\,P_\infty ) \int _{Q_1}^{-\infty} dQ/P(Q|g) B^{1/2}(Q),
\end{align}
whereas  $\tau_{**}'$ is obtained by changing the sign in the last term of the above expression; $P_\infty$ is the value of the momentum for large negative $Q$ (note that Im~$P$ does not change sign in the range $Q_1 > Q$ for the considered range of $g$). 

Near these poles we find that
\begin{align}
\label{eq:pole2}
&Q\approx \frac{i}{2(t - \tau_{**})}, \quad P \approx -iQ, \quad  |t - \tau_{**}|\ll 1,\\
&Q\approx \frac{-i}{2(t - \tau_{**}')}, \quad P \approx iQ, \quad  |t - \tau_{**}'|\ll 1.
\end{align}

In the range of $g$ where $g(Q,P)$ has one well, to find the poles $\tau_{**}$ and $\tau_{**}'$ one should move from $Q_3$ to $Q\to -\infty$ along the real $Q$-axis with Im~$P>0$ and Im~$P<0$, respectively. If the points $Q_\pm$ are on the real $Q$-axis, moving between them will contribute a real part $\tau_p\1/2$ to $\tau_{**}$ and $\tau_{**}'$.

\begin{figure}[h]
	\includegraphics[width=1\linewidth]{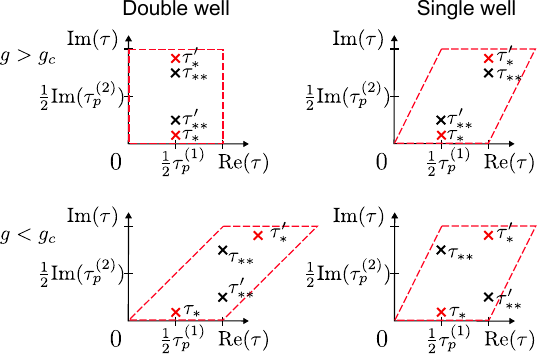}
	\caption{Position of the poles of $Q(\tau)$ and $P(\tau)$ for $Q(0) = Q_3$ and $P(0) = 0$. The dashed lines show the parallelogram of periods, with the real period $\tau_p^{(1)}$ and the complex period $\tau_p^{(2)}$.}
	\label{fig:complex_poles}
\end{figure}

The positions of the poles together with the parallelograms of periods $\tau_p^{(1)}$ and $\tau_p^{(2)}$ are shown in Fig.~(\ref{fig:complex_poles}). The parallelograms have the same shape for a shallow well in the range $g>g_\mathrm{min}^\mathrm{shallow}$. Indeed, one can make a transition between the trajectories in the wells by changing time by $\tau_p\2/2$, which is the time it takes to move from the turning point in the deeper well $Q_3$ to the turning point in the shallower well $Q_2$, for the chosen $\alpha_d>0$. The positions of the poles and the forms of the parallelogram of periods are different for the single- and double-well cases and for $g>g_c$ and $g<g_c$.

\subsubsection{Action as a function of imaginary time}
\label{subsubsec:action}

An extra insight into the dynamics of the system can be gained by looking at the action for a classical trajectory evaluated as a function of the complex time $\tau$,
\begin{align}
\label{eq:action_tau}
I(g|\tau) = \frac{1}{2\pi}\int_{\tau}^{\tau+\tau_p\1} dt \,P(t) \dot Q(t)\,.
\end{align}
The integral can be thought of as starting at a point with complex $\tau$ on the boundary of the parallelogram of periods in Fig.~\ref{fig:complex_poles} and moving parallel to the real-time axis to the next boundary. Thus calculated action $I(g|\tau)$ is shown in Fig.~\ref{fig:topology}~(d)-(f). Since $Q(t+\tau)$ and $P(t+\tau)$ are in general complex  for imaginary $\tau$, it does not correspond to the areas of the trajectories in  Fig.~\ref{fig:topology}~(a)-(c). Note that plotted in  Fig.~\ref{fig:topology}~(a)-(c) are the trajectories in the variables $\bigl(\arctan[\mathrm{Re}~Q], \arctan[\mathrm{Re}~P]\bigr)$, which allowed us to take into account that $|Q|$ and $|P|$ diverge as $\tau$ approaches $\tau_*$ or $\tau_{**}'$. 

The functions $P(t), Q(t)$ do not have branching points. Therefore, based on the picture of the poles in Fig.~\ref{fig:complex_poles}, one would expect that $I(g|\tau)$ is equal to the action $ I(g|0)$ for the intrawell classical orbit for a given $g$ as long as Im~$\tau <$~Im~$\tau_*$, in which case the integration contour can be shifted to the real axis Im~$\tau = 0$. Once Im~$\tau$ goes over the lowest pole, Im~$\tau_*$, the action should decrease by the corresponding residue $\oint P\,dQ/2\pi=(\mu + \alpha)/2$. We used here that near this pole $P\approx -i(Q-1) + i(\mu + \alpha_d)/2Q$.   Once Im~$\tau$ goes over the next pole, Im~$\tau_{**}'$, another residue contributes to the integral. Here  $P\approx i(Q +1) - i(\mu - \alpha_d)/2Q$. Thus the overall change of the action $\oint P\,dQ/2\pi$ by going over both residues is $- \alpha_d$. This is also the difference in the actions in the wells $\sigma=-1$ and $\sigma=1$, since the action for the $\sigma=-1$-well is determined by the trajectory $Q(t + \tau_p\2/2), P(t+\tau_p\2/2)$ with real $t$, while there are no poles along the imaginary-time axis between $\tau'_{**}$ and Im~$\tau_p\2/2$.


\section{Quantum heating}
\label{sec:activation}

The focus of this paper is the effect of the dynamical bias on the phase-flip rate, i.e., the rate of switching between the wells of $g(Q,P)$. Related to this is the problem of the distribution over the intrawell states. The distribution is formed as a result of random walk over the intrawell Floquet states. Such walk is a consequence of the coupling of the oscillator to a thermal bath. In the absence of modulation,  for $T=0$ the oscillator makes downward transitions between its energy levels (separated by $\approx\hbar\omega_0$)  with emission of excitations into the bath. In the presence of modulation, these transitions correspond to upward and downward  transitions between the quasienergy states, since these states are linear combinations of the states of the undriven oscillator. The outcome is occupation of excited intrawell states in the stationary regime, the effect called quantum heating \cite{Dykman2012a,Ong2013}. 


\subsection{The rates of transitions between the Floquet states}
\label{subsec:transition_rates}

In the analysis of the dissipative dynamics of the modulated oscillator, we will assume  that the well-understood conditions of the applicability of the Markovian master equation  are met, cf.~\cite{Kamenev2023}. We will further assume that the transition rates are small compared to the spacing between the quasienergy levels. Then  the dynamics of the oscillator can be described  by a balance  equation for the populations $\rho_n$ of the Floquet (quasienergy) states,
\begin{align}
\label{eq:balance}
	\dot \rho_n &= -\sum_{m} (W_{n\, n+m}\rho_n - W_{n+m\, n} \rho_{n+m}),
\end{align}
where $n$ enumerates the states. The explicit form of the rates $W_{n'n}$ of  $n'\to n$  transitions depends on the relaxation mechanism. We will consider it for the most commonly encountered mechanism where the coupling  of the oscillator to the thermal bath is linear in $\hat a, \hat a^\dagger$. Then, in the rotating frame, relaxation is described by the superoperator $\mathcal{L}[\rho] = 2\kappa[\hat a \rho \hat a^\dagger  - \frac{1}{2}(\hat a^\dagger \hat a \rho + \rho \hat a^\dagger \hat a) ]$. The parameter $\kappa$ is the dimensionless  oscillator decay rate; it is the coefficient of viscous friction force in the classical limit scaled by the factor $2m_0\omega_p/F_p$. Relaxation in dimensional time is often described by the same operator $\mathcal{L}[\rho]$ with $\kappa$ replaced by $\kappa/2$. 

The expression for $\mathcal{L}$ is written in the low-temperature limit, where the thermal occupation number of the oscillator $\bar n =[\exp{\hbar \omega_p/2k_BT} -1]^{-1}$ can be set equal to zero. From this expression, the transition rate is
\begin{align}
\label{eq:transition_rate}
W_{nn'} = 2\kappa \abs{\braket{n'|\hat{a}|n}}^2  \,.
\end{align}

In the semiclassical case $\lambda\ll 1$ the matrix elements $ \braket{n'|\hat a |n}$ are expressed in terms of the Fourier components $a_m(g)$ of the functions $a(\tau,g) = (2\lambda)^{-1/2}[P(\tau,g) - iQ(\tau,g)] $. Here $Q(\tau,g)$ and $P(\tau,g)$ are the coordinate and momentum as functions of time for a given $g$; these functions are periodic with period $\tau_p\1(g)=2\pi/\omega(g)$,
\begin{align}
\label{eq:Fourier_components}
&\braket{n+m|\hat a|n} \approx a_m(g_n), \nonumber\\
&a_m(g)=\frac{1}{2\pi}\int_0^{2\pi}d\phi\,a(\tau,g)e^{-im\phi}, \quad \phi = \omega(g)\tau
\end{align}

The matrix elements can be evaluated explicitly making use of the double periodicity of $Q(\tau,g)$ and $P(\tau,g)$. We start with the matrix elements for the well at $Q>0$. One can then relate the integral  in Eq.~(\ref{eq:Fourier_components}) to a contour integral.  In the complex $\tau$-plane the contour of choice, which we denote by $\mathcal{C}$,  is a parallelogram that goes from $0$ to $\tau_p\1$, then to $ \tau_p\1 + \tau_p\2$, then to $\tau_p\2$, and then back to $0$. In terms of the phase $\phi = \omega(g)\tau$, the contour $\mathcal{C}$ goes along the loop $0 \to 2\pi \to 2\pi+ \omega(g)\tau_p\2 \to \omega(g)\tau_p\2\to 0$. Because of the periodicity $a(\tau,g) = a(\tau+\tau_p\1,g) = a(\tau+\tau_p\2,g)$, we have
\begin{align}
&a_m(g) = \frac{1}{2\pi} \left(1 - e^{-im\phi_p\2}\right)^{-1}\oint_\mathcal{C} d\phi \, a(\tau,g) e^{-im \phi}\,,\nonumber\\
&\phi_p\2 = \omega(g)\tau_p\2.
\end{align}

As described in Sec.~\ref{subsec:poles}, the parallelogram $\mathcal{C}$ contains four poles of $Q(\tau,g)$ and $P(\tau,g)$. However, from Eqs.~(\ref{eq:pole1}) and (\ref{eq:pole2}), only $\tau_\ast$ and $\tau_{\ast\ast}$ are poles of $a(\tau,g)$. With the account taken of these equations, 
\begin{align}
	\label{eq:matrix_elements}
	a_m(g) &= - \frac{i\omega(g)}{\sqrt{2\lambda}}\frac{\exp(-im\phi_\ast) - \exp(-im\phi_{\ast\ast})}{1- \exp(-im\phi_p^{(2)})}\,,
\end{align}
where $\phi_\ast = \omega(g) \tau_\ast$, $\phi_{\ast\ast} = \omega(g) \tau_{\ast\ast}$. 

The absolute value of the matrix elements decays exponentially for large $|m|$. The exponent is different for $m\gtrless 0$. We have
\begin{align}
	a_m (g) &\propto \mathrm{exp}(-m\Im(\phi_p^{(2)} - \phi_{\ast\ast})),&  m&\gg 1\,,\nonumber\\
	a_m (g) &\propto \mathrm{exp}(-|m|\Im \phi_\ast), & -m&\gg 1 \,.
\end{align}

The matrix elements $a_m(g)$ for the $\sigma=-1$-well in the range $g>g_\mathrm{min}^\mathrm{shallow}$ can be found by noting that, in this range of $g$,  one can switch from a trajectory in one well to the trajectory in the other well with the same $g$ by shifting  time by $(\tau_p\1+\tau_p^{(2)})/2$, which is the time to move along the Hamiltonian trajectory from $Q_3$ to $Q_1$. The contour of integration $\mathcal {C}$ has to be shifted appropriately. As a result, the matrix elements $a_m(g)$ for the $\sigma=-1$-well are then given by Eq.~(\ref{eq:matrix_elements}) with $\phi_\ast$ and $\phi_{\ast\ast}$ replaced by $\tilde \phi_\ast = \phi_{\ast\ast}-\omega(g)\tau_p^{(2)}/2$ and $\tilde \phi_{\ast\ast} = \phi_\ast + \omega(g) \tau_p^{(2)}/2$, respectively.

For $\alpha_d <0$, where the $\sigma=-1$-well is the deeper well of $g(Q,P)$, the expressions for the Fourier components have to be interchanged. As seen from Eqs.~(\ref{eq:transition_rate}) and (\ref{eq:Fourier_components}), these expressions give the transition rates $W_{n\,n+m}$ in the explicit form. The rates $W_{n\,n+m}$ fall off exponentially with the increasing $|m|$.

For $|m\alpha_d|\ll 1$ one can expand the result for the matrix elements and find a simple analytic expression for the linear in $\alpha_d$  correction to $a_m(g)$. It agrees numerically with the result found in \cite{Boness2024a}, which involved summation over the matrix elements for $\alpha_d=0$.

\subsection{Intrawell probability distribution in the semiclassical limit}
\label{subsec:tail}

It follows from the explicit expressions for the transition rates that the oscillator most likely makes transitions between the Floquet states $\ket{n}$  toward the minimum of the well  of $g(Q,P)$ it currently occupies. However, states remote from the minimum can be also reached, albeit with small probabilities. In the semiclassical range $\lambda\ll 1$, where there are many quasienergy levels inside the wells of $g(Q,P)$, these probabilities can be found using the real-time-instanton approach, see \cite{Dykman2012a,Kamenev2023} and references therein. The approach relies, in part, on the time scale separation. Over the time $\sim \kappa^{-1}$ there is formed a quasistationary distribution over the intrawell states. Switching between the wells is an unlikely event, as it involves reaching highly excited intrawell states and going over the barrier that separates the wells.  The switching rates $W_\mathrm{sw}$ are exponentially smaller than $\kappa$, see Sec.~\ref{sec:switching_rates}.

Following this approach, we seek the quasisationary intrawell distribution over the states in the form
\begin{align}
	\rho_n = e^{-R(g_n)/\lambda}\,.
\end{align}
This expression reminds the Boltzmann distribution. The parameter $\lambda$, which characterizes quantum fluctuations, plays the role of temperature. Assuming that $R(g)$ is a smooth function of $g$ and using that $W_{n\,n+m}\approx W_{n-m \, n}$ in the semiclassical range, we write Eq.~(\ref{eq:balance}) as
\begin{align}
 \label{eq_Jacobi}
	\partial_t R\equiv \frac{\partial R}{\partial t} &= - \mathcal{H}(I, \partial_I R)\,,  \nonumber\\
	\mathcal{H}(I, p_I) &= \lambda \sum_m W_{n-m\, n} (e^{m p_I} -1)\, .
\end{align}
This equation has the form of a Hamilton–Jacobi equation for an auxiliary system with the coordinate $I$, momentum $p_I$, and Hamiltonian $\mathcal{H}$, whereas $R$ is the action variable. The dependence of $\mathcal{H}$ on $I$ comes from the transition rates $W_{n-m\,n}$, which have to be evaluated for $n$ given by the Bohr-Sommerfeld condition  $I=\lambda \bigl(n+(1/2)\bigr)$.

The Hamiltonian equations of motion for $I$ and $p_I$ are 
\begin{align}
\label{eq:instanton_eqs}
	\dot{I} &= \frac{\partial \mathcal{H}}{\partial p_I}\,, & \dot{p}_I &= - \frac{\partial \mathcal{H}}{\partial I}\,.
\end{align}
The quasistationary distribution for $\rho_n$ is given by the stationary solution of Eq.~(\ref{eq_Jacobi}), i.e., by the solution of the equations of motion (\ref{eq:instanton_eqs}) with $\mathcal{H}=0$. 

A trivial solution of the equation $\mathcal H(I, p_I) = 0$ is found by setting $p_I=0$. In this case, with the account taken of  Eq.~(\ref{eq:transition_rate}) and using the method \cite{Peano2014} one can show that the equation for the action variable (\ref{eq:instanton_eqs}) takes the form $\dot{I} = - 2\kappa I$. This equation describes relaxation of the system toward the bottom of the initially occupied well of $g(Q,P)$ in the neglect of quantum fluctuations.

The solution of the Hamiltonian equations (\ref{eq:instanton_eqs}) that describes the distribution $\rho_n$, i.e., the  real-time-instanton solution, corresponds to a trajectory $I(t)$ that goes away from the bottom of the well.  It can be  found by noting that the transition rates obey a detailed balance condition 
\[W_{nn'}W_{n'n''}W_{n''n} =  W_{nn''}W_{n''n'} W_{n'n},\]
since
\begin{align}
	\frac{W_{n-m\, n}}{W_{n+m\, n}} &= e^{2m \Im\,(\phi_{\ast\ast} + \phi_\ast - \phi_p^{(2)})}\,, \label{eq:detailed_balance}
\end{align}
as seen from Eq.~(\ref{eq:matrix_elements}). Note, that Eq.~(\ref{eq:detailed_balance}) holds for both wells. The condition $\mathcal{H}(I,p_I) = 0$ is then fulfilled for 
\[p_I = 2\Im\,(\phi_p^{(2)} - \phi_\ast - \phi_{\ast\ast})\,,\] 
which gives
\begin{align}
\label{eq:instanton_explicit}
 \dot I = 2\kappa I.
 \end{align}
Thus, the  instanton trajectory $I(t)$ is simply given by the time reversed fluctuation-free path. The trajectory  (\ref{eq:instanton_explicit}) goes from the bottom of the occupied  well of $g(Q,P)$ up to larger $g$. We note that, in general, out-of equilibrium systems do not have time-reversal symmetry. Beyond a narrow range of $\bar n$ that shrinks to zero for $\lambda\to 0$, the optimal trajectory $I(t)$ differs from the time-reversed fluctuation-free path.

In the considered case $\bar n = 0$ it follows from the above expressions that
\begin{align}
\label{eq:R_prime}
	R'(g) \equiv \partial_g R = \omega^{-1}(g) p_I =  2 \Im(\tau_p^{(2)} - \tau_\ast - \tau_{\ast\ast})\,
\end{align}
independent of the well. Thus the function $R(g)$, which describes the tale of the distribution over the intrawell states in the considered case $\bar n=0$, is  given by an integral of a simple explicit expression. One can show that $\Im\,(\tau_*+\tau_{**})  <\Im\, \tau_p\2$, see Appendix~\ref{sec:append_relation}. Therefore, $R'>0$, and the quasistationary state population  
\begin{align}
\label{eq:state_populations_explicit}
\rho_n = C_\mathrm{w}\exp\left[-\int_{g_\mathrm{min}}^{g_n} dg R'(g)/\lambda
\right]
\end{align}
monotonically falls off with the increasing distance $g_n-g_\mathrm{min}$ from the bottom $g_\mathrm{min}$ of the occupied well of $g(Q,P)$. The parameter $C_\mathrm{w}$ here is the normalization constant that depends on the occupation of the well.

\begin{figure}[h]
	\includegraphics[width=1\linewidth]{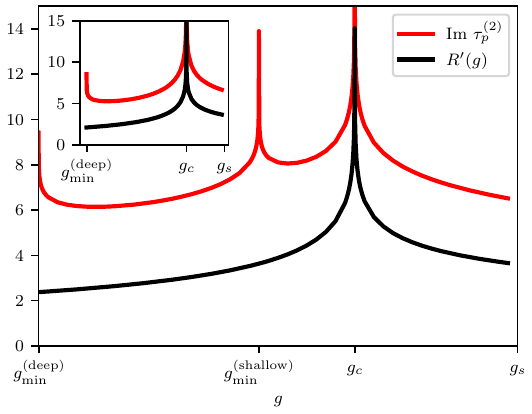}
	\caption{The imaginary part of the complex vibration period $\tau_p\2$ and $R'(g)$ for the scaled detuning of the  modulation frequency $\mu=0.2$ and the scaled amplitude of the biasing field $\alpha_d = 0.1$. The function $\mathrm{Im}~\tau_p^{(2)}(g) $ diverges logarithmically at $g_\mathrm{min}^\mathrm{deep}$, $g_\mathrm{min}^\mathrm{shallow}$, and $g_c$, while $R'(g)$ diverges  only at $g_c$. The inset refers to $\mu=\alpha_d=0.2$, in which case $g_c = g_\mathrm{min}^\mathrm{shallow}$.}
	\label{fig:rprime}
\end{figure}

In Fig.~\ref{fig:rprime} we show $R'(g)$ together with the imaginary part of the complex period $\mathrm{Im}~\tau_p^{(2)}$. The function  $\mathrm{Im}~\tau_p^{(2)}$ diverges logarithmically at $g_c$, as discussed above. It also diverges logarithmically near the bottoms of the wells $g_\mathrm{min}^\mathrm{deep}$ and  $g_\mathrm{min}^\mathrm{shallow}$, see Appendix~\ref{sec:append_minima}. In contrast, $R'(g)$ only diverges at $g_c$. Near the bottoms of the wells the distribution $\rho_n$ can be found analytically, since Eqs.~(\ref{eq:eom}) for the dynamical variables $Q,P$ can be linearized, see Appendix~\ref{sec:append_minima}.


\section{Quantum activation}
\label{sec:switching_rates}

Equation (\ref{eq:state_populations_explicit}) shows that, even for zero temperature, the oscillator occupies states far from the bottom of the wells of $g(Q,P)$, including the states near the boundary that separates the wells.
If the oscillator was initially prepared in one of the wells, once it reaches the near-boundary states, it will switch to another well with probability $\sim 1/2$. This is reminiscent of thermal activation where a classical system switches between potential wells due to thermal fluctuations. Since in the considered case the switching is due to quantum fluctuations, it is called quantum activation \cite{Marthaler2006,Kamenev2023}. 

The minimal value of $g(Q,P)$ at the interwell boundary is the saddle-point value $g_s$. Therefore the switching rate is 
\begin{align}
\label{eq:switching_rate}
	W_\mathrm{sw}(\sigma) &= C_\mathrm{sw}(\sigma) \times \exp(-R_A(\sigma)/\lambda)\,,\nonumber \\
	R_A(\sigma) &= \int_{g_\mathrm{min}(\sigma)}^{g_s} R'(g) d g\,,
\end{align}
where $\sigma= \pm 1$ enumerates the wells; the prefactor $C_\mathrm{sw}(\sigma)$ is of the order of the decay rate (of the order of $\kappa$, in the dimensionless time). The parameter $\lambda$ characterizes the intensity of quantum fluctuations, and therefore, by analogy with thermal activation where the intensity of fluctuations is characterized by temperature, $R_A$ can be called  the quantum activation energy.

\begin{figure}[h]
	\includegraphics[width=1\linewidth]{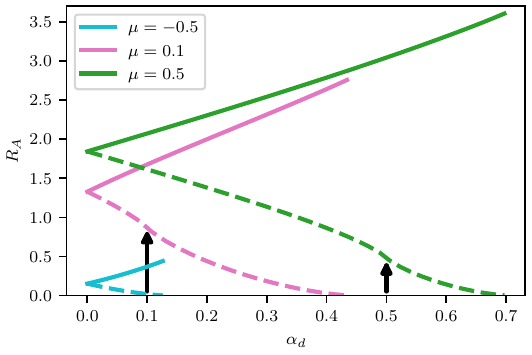}
	\caption{The dependence of the activation energies $R_A$ for switching from the deep (solid lines) and shallow (dashed lines) wells on the bias $\alpha_d$. The arrows indicate the values of $\alpha_d$ where the minimum of the shallow well coincides with $g_c$. The results refer to $\mu=-0.5, 0.1$, and $0.5$.}
	\label{fig:activation}
\end{figure}

We show the activation energies $R_A$  in Fig.~\ref{fig:activation}. For the deeper and shallower wells, $R_A$  monotonically increases and monotonically decreases with the increasing  $\alpha_d$, as do also the depths of the wells. The activation energy of switching from the shallow well $R_A^\mathrm{shallow}$ goes to zero as $|\alpha_d|$ approaches the bifurcational value $2[(\mu + 1)/3]^{3/2}$ where the well disappears.  We note that, since $R'(g)$ is independent of the well, the difference in the values of $R_A$ is given by 
\begin{align}
\label{eq:Delta_R_A}
\Delta R_A = \int_{g_\mathrm{min}^\mathrm{deep}}^{g_\mathrm{min}^\mathrm{shallow}} R'(g) \,dg,
\end{align}
i.e., by the integral of $R'(g)$ over the difference in the minimal values  $g_\mathrm{min}^\mathrm{shallow}$ and $g_\mathrm{min}^\mathrm{deep}$ of $g(Q,P)$ in the shallow and deep wells.  
 
The difference in $R_A$ between the wells leads to an exponentially large difference in the switching rates.  Respectively, the stationary populations of the wells $w(\sigma)$, which are determined by these rates, are also exponentially different,
\[w(1)/w(-1) \propto \exp\left[\bigl(R_A(1) - R_A(-1)\bigr)/\lambda\right].\] 

Along with interwell switching via quantum activation, the oscillator can also switch via quantum tunneling. Where there is no bias, $\alpha_d=0$, the wells of $g(Q,P)$ are symmetric and interwell tunneling  is resonant for all states. Bias  shifts quasienergy levels of different wells out of resonance, generally. Since the action between the two wells differs by $\delta I = |\alpha_d|$, only for 
\begin{align}
\label{eq:interwell_resonance}
\alpha_d = m\lambda, \qquad m=\pm 1, \pm 2, ...
\end{align}
are the semiclassical states in the shallow well of $g(Q,P)$ in resonance with the states in the deeper well. This relation coincides with the relation used in \cite{deAlbornoz2024} to describe the experimental results.  For the lowest quasienergy level in the shallow well $I=\lambda/2$, and therefore Eq.~(\ref{eq:interwell_resonance}) is the condition on the difference of the well depths. Importantly, since the vibration frequencies in the wells are the same for the same $g$, all levels come in resonance at a time. 

Even in the resonant case, switching via quantum activation is exponentially more likely than via tunneling in the important case where the tunnel splitting is small compared to the dissipation-induced broadening of the intrawell quasienergy levels. This can be shown by extending the arguments given in \cite{Marthaler2006} for the symmetric case.  There it was shown that escape via quantum activation dominates if Im~$\tau_p\2 >R'(g)$. This condition is seen from Eqs.~(\ref{eq:tau*}), (\ref{eq:tau**}), and (\ref{eq:R_prime}) to hold also in the asymmetric case we consider here. We note that, in contrast to quantum activation, the tunneling amplitude is not proportional to the relaxation rate, and therefore tunneling can become important for very small $\kappa$, in particular, for $\kappa$ small compared to the tunnel splitting.

\subsection{The singularity of $R'$}
\label{subsec:R_prime_singularity}

The logarithmic divergence of $R'(g)$ at $g=g_c$ indicates that the  eikonal approximation used to  obtain Eq.~(\ref{eq_Jacobi}) breaks down for $|g-g_c|\ll 1$. However, the divergence is integrable. The contribution of the range of small  $|g-g_c|$ to the effective activation energy $R_A$ is small. This was confirmed by directly solving the Schr\"odinger equation $\hat g \ket{n} = g_n\ket{n}$ for the case $\alpha_d=0$ \cite{Marthaler2006}. 

One can understand it alternatively from  the analysis of the matrix elements $\braket{n+m|\hat a |n}\approx a_m(g_n)$. These matrix elements determine the probabilities of transitions $\ket{n}\to \ket{n+m}$; note that transitions with $m>0$ are the transitions up the $g$-axis; they ultimately lead to interwell switching. In the double-well regime, the factor $\exp(-im\tau_p\2(g)/2)$, which determines $a_m(g)$, changes sign as $g$ goes across $g_c$  for $m=2k+1$ and $k\geq 0$. From Eq.~(\ref{eq:matrix_elements}), so does also $i a_m(g)$. Therefore some of the matrix elements $\braket{m+n|\hat a|n}$, which generally contain a small quantum correction to $a_m(g_n)$, may become equal to zero for $g_n$ close to $g_c$. Then there would be no transitions up by the proper $m$ levels from the corresponding states. However, generically, this requires fine tuning and may happen to one state. Even in this case there would be transitions up along the $g$-axis from other states. This is why the calculation of $R_A$ in the eikonal approximation is justified.

A special case is where $g_c$ coincides with the bottom of a well, so that in this well there are no states $\ket{n}$ with  $g_n< g_c$. This happens for  $\alpha_d=\pm \mu$. Near the bottom of a well the Hamiltonian $\hat g(Q,P)$ is approximated by the Hamiltonian of a harmonic oscillator, cf. Appendix~\ref{sec:append_minima}. Respectively, the matrix elements $\braket{m|\hat a|n} \propto \delta_{m,n\pm 1}$ are limited to transitions between the nearest states.  However, for 	$g_c=g_\mathrm{min}$ we find that $\braket{1|\hat a|0} = 0$, where $\ket{0}$ is the ground (lowest-$g$) state in the well. This is in agreement with the semiclassical result $R'(g_c)\to \infty$.  As a consequence, if the oscillator is in the ground intrawell state, it will stay there. In the symmetric case, $\alpha_d=0$, and for $\mu=0$ this was indicated in Ref.~\cite{Marthaler2006} and was recently seen in Ref.~\cite{Frattini2024} as an extremely slow switching for small $\bar n$. Generally, one may still expect quantum heating in this case due to transitions to the excited states of $\hat g$ which, however, have exponentially small amplitudes.

For $0<\mu <2$ the condition $g_c = g_\mathrm{min}$ can hold for a shallower well. The values of $\alpha_d$ where it happens are marked by the arrows in Fig.~\ref{fig:activation}. If the system is initially prepared in this well, it will be trapped there, even though the well is expected to be less populated. Such trapping  provides an insight into  one of the  features of the stationary distribution of the oscillator found  in \cite{Roberts2020}.

We note that, generally, the oscillator can escape from the shallow well via tunneling unless  tunneling is suppressed by the quantum interference in the classically inaccessible region; such suppression was predicted \cite{Marthaler2007} and recently observed \cite{Venkatraman2024} in the absence of bias. We also note that  corrections from finite $\bar{n}$ as well as dephasing change the transition rates $W_{n~n+m}$ and facilitate escape of the oscillator \cite{Marthaler2006,Boness2024a}. Dephasing due to slow frequency fluctuations, that result from $1/f$-type noise,  can be particularly important in this respect as such fluctuations drive the system away from the parameter values where $g_c = g_\mathrm{min}$; the analysis of this effect is beyond the scope of the present paper.


\subsection{Logarithmic susceptibility}
The switching rates may change dramatically already for a weak bias, $|\alpha_d| \ll 1$, provided  $|\alpha_d| \gg \lambda$ \cite{Boness2024a}. For $|\alpha_d|\ll 1$ the activation energies become
\begin{align}
\label{eq:R_A_linear}
	R_A(\sigma) &\approx R_A^{(0)} +\sigma \alpha_d R_A^{(1)}\,.
\end{align}
Here, $R_A^{(0)}=R_A^{(0)}(\pm 1)$ is the activation energy for the symmetric case, $\alpha_d = 0$, where both minima are located at $Q=Q_\mathrm{min} = \sigma (\mu+1)^{1/2}$ and $g=g_\mathrm{min}^{(0)}=-(\mu+1)^2/4$. The linearity of the correction to $R_A$ in $\alpha_d$ for small $|\alpha_d|$ is a consequence of the explicit expressions (\ref{eq:R_prime}) and (\ref{eq:switching_rate}). As seen from these expressions, $R'(g)$ is an even function of $\alpha_d$: the change $\alpha_d\to -\alpha_d, Q\to -Q, P\to -P$ does not change the Hamiltonian $\hat g(Q,P)$ and the equations of motion (\ref{eq:eom}).  Therefore the change $\alpha_d\to -\alpha_d$ does not change the values of $\tau_p\2, \tau_*$, and $\tau_{**}$ in Eq. ~(\ref{eq:R_prime}) for $R'(g)$. Hence, the expansion of $R'(g)$ in $\alpha_d$ starts with $\alpha_d^2$. We took into account here that  $R'(g)$ is a smooth function of $g$ away from $g_c$. To consider the effect of the vicinity of $g_c$, we note that, from Eq.~(\ref{eq:critical_g}),  the correction to $g_c$ is quadratic in $\alpha_d$ for $\alpha_d\to 0$.
Therefore the correction to Im~$\tau_p\2$, and thus to $R'(g)$, is also quadratic in $\alpha_d$ for $g$ close to $g_c$; see, however, the discussion of the case $\mu=0$ below. 

On the other hand, the correction to $g_\mathrm{min}$ in Eq.~(\ref{eq:switching_rate}) is linear in $\alpha_d$. For the well $\sigma$ we find from the expression for $g(Q,P)$ that 
\[g_\mathrm{min}(\sigma)-g_\mathrm{min}^{(0)} \approx -\sigma\alpha_d (\mu+1)^{1/2}.
\]
It is the shift of $g_\mathrm{min}$ that leads to a linear in $\alpha_d$ term in Eq.~(\ref{eq:R_A_linear}). From Eq.~(\ref{eq:switching_rate}) we have
\begin{align}
\label{eq:log_sus}
	R_A^{(1)} = &(\sigma/\alpha_d)\int_{g_\mathrm{min}(\sigma)}^{g_\mathrm{min}^{(0)}}dg R'(g)\nonumber\\
&	\approx (\mu +1)^{1/2} R'^{(0)}(g_\mathrm{min}^{(0)})\,,
\end{align}
where $R'^{(0)}(g_\mathrm{min}^{(0)})$ is the value of $R'(g)$ evaluated at the minimum of $g(Q,P)$ for $\alpha_d = 0$. This value was found in  \cite{Marthaler2006}. Using it we obtain 
\begin{align}
\label{eq:R_A_1_explicit}
	R_A^{(1)} &= \frac{1}{2} \log(\frac{\mu+2 + 2 \sqrt{1+\mu}}{\mu+2 - 2 \sqrt{1+\mu}})\,.
\end{align}
\begin{figure}[h]
	\includegraphics[width=1\linewidth]{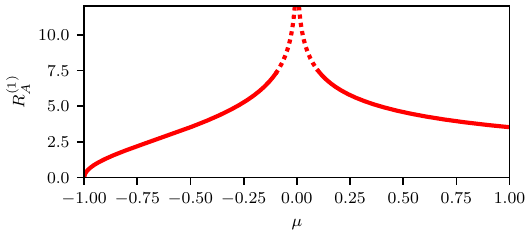}
	\caption{Logarithmic susceptibility $R_A^{(1)}$ as a function of the scaled frequency detuning of the parametric drive $\mu$. The function  $R_A^{(1)}$ gives the correction to the switching activation energy that is linear in  the scaled amplitude $\alpha_d$ of the symmetry-breaking field, see Eq.~(\ref{eq:R_A_linear}). For $\mu\to 0$ the perturbation theory, which leads to Eq.~(\ref{eq:R_A_linear}), becomes inapplicable. In this region the result is shown by a dashed line.}
	\label{fig:susceptibility}
\end{figure}

Equation (\ref{eq:R_A_linear})  shows that the correction to the logarithm of the switching rate is linear in the bias field. Therefore the coefficient $R_A^{(1)}$ can be called the logarithmic susceptibility. For $\bar n \gg \lambda$ it was found   in Ref.~\cite{Boness2024a}. 

We show the result for the correction to the activation energy for $\bar n = 0$ in Fig.~(\ref{fig:susceptibility}). This correction monotonically increases as $|\mu|$ decreases. 

For $|\mu|\ll 1$ the perturbation theory leading to Eq.~(\ref{eq:R_A_linear}) becomes inapplicable. In this range, for small $|\mu|$ and small $|\alpha_d|$ we have, from Eq.~(\ref{eq:R_prime_min}), $R'\bigl(g_\mathrm{min}(\sigma)\bigr) \approx \log (4/|\mu + \sigma\alpha_d|)$. Respectively, we eliminated the range of small $|\mu|$ in Fig.~\ref{fig:susceptibility}. Physically, for $\mu+ \sigma\alpha_d = 0$ the system is localized in the ground intrawell state. The rate of transitions from this state $W_{01}$ can be shown to be $\propto |\mu+ \sigma\alpha_d|^2$ for small $ |\mu+ \sigma\alpha_d|$.  Again, the localization is destroyed and the correction to $R_A$ becomes linear in $\alpha_d$ for not too small $\bar n$ and in the presence of dephasing. 

\subsection{Prebifurcation regime}
\label{subsec:bifurcation}

The problem of switching from a shallow well simplifies near the bifurcational value $\alpha_B$ of the bias $\alpha_d$, where the well disappears, $|\alpha_B|= 2[(1+\mu)/3]^{3/2}$. The minimum of the well merges with the saddle point of $g(Q,P)$ for $Q=Q_B, P=0$, where $Q_B = - [(1+\mu)/3]^{1/2}\sgn\,\alpha_d$. At the bifurcation point $g(Q_B,0) = (1+\mu)^2/12$. For concreteness we will assume that $\alpha_d$ is close to $|\alpha_B|>0$. In this case the function $g(Q,P)$ has two wells for $\delta\alpha_d = \alpha_d - |\alpha_B| <0$. It is easy to show that, in the vicinity of the shallow well, to the leading order in $\delta\alpha_d$,
\begin{align}
\label{eq:g_bif}
g(Q,P) - g_B \approx \frac{2-\mu}{3}P^2  +Q_B\delta Q^3 - \delta\alpha_d \delta Q - Q_B\delta\alpha_d,       
\end{align}
where $\delta Q = Q-Q_B, |\delta Q| \ll \sqrt{(1+\mu)/3}$. This maps the problem of switching from the shallow well on the problem of switching of a mechanical particle with effective mass  $3/2(2-\mu)$ from a potential of the form of the cubic parabola.

At the bifurcation point the frequency $\omega(g)$ goes to zero.  The dynamics of the system is different if, for the considered bias $\alpha_d \approx \alpha_B$ this frequency is large or small compared to the decay rate $\kappa$. We call ``preburcation regime'' the parameter range where $|\alpha_d - \alpha_B|\ll 1$ yet the vibrations about the minimum of the shallow well of $g(Q,P)$ are underdamped, $\omega(g_\mathrm{min}^\mathrm{shallow}) \gg \kappa$. In this regime there apply the general expression (\ref{eq:switching_rate}) for the switching rate. 

The analysis is simplified by the fact that  $\omega(g_\mathrm{min}^\mathrm{shallow}) \approx 2[(2-\mu)]^{1/2}(|\delta\alpha_d Q_B|/3)^{1/4}$ is small in the prebifurcation regime. Therefore one can use the approach \cite{Marthaler2006,Boness2024a} to write 
\begin{align}
\label{eq:prebif_answer}
&	R'(g) = 2 \frac{M(g)}{N(g)}\,, \qquad	M(g) = \iint_{A(g)} d{}{Q} d{}{P}\,,
	\nonumber\\
&	N(g) = \frac{1}{2} \iint_{A(g)} d{}{Q} d{}{P} \,(\partial_Q^2 + \partial_P^2) g(Q,P)\,,
\end{align}
where the integrals run over the interior $A(g)$ of the well limited by the contour $g (Q,P) = g$. 
From Eq. ~(\ref{eq:g_bif}) we see that $M(g)/N(g) = 3/(2-\mu)$, to the leading order in $\delta\alpha_d$. Using Eq.~(\ref{eq:g_bif}) to find the depth of the shallow well,  we obtain for  the activation energy for switching from the shallow well near the bifurcation point the expression
\begin{align}
\label{eq:shallow_bif}
R_A^\mathrm{shallow} = \frac{8}{2-\mu}|\delta\alpha|^{3/2}/[3(1+\mu)]^{1/4}.
\end{align}
This expression displays the characteristic scaling of the switching rate $|\alpha_d|^{3/2}$ with the distance $|\delta\alpha_d|$ to a saddle-node bifurcation point, cf. \cite{Dykman2012a} and references therein.


\section{Conclusion}
\label{sec:conclusions}

In this paper we studied the rates of  switching between the states of parametrically excited vibrations of a quantum oscillator. The states have opposite phases, and the switching is a phase flip. We  considered the effect of an extra field at half the frequency of the parametric modulation. Such field provides a dynamical bias that lifts the symmetry with respect to time translation by the modulation period, making the switching rates different for the states with different  phases. Our analysis refers to the coupling to a low-temperature thermal reservoir, where the thermal occupation number of the oscillator in the absence of modulation can be set  equal to zero. We assume that the coupling is weak, so that the dissipation rate is small.

As a function of the coordinate and momentum in the rotating frame, the Floquet Hamiltonian of the oscillator has a double-well form, with the minima corresponding to the states with opposite phases. The wells are asymmetric because of the dynamical bias. The interwell switching results from quantum activation, i.e., from transitions over the barrier that separates the wells. It is due to unavoidable quantum fluctuations that come along with dissipation in the presence of modulation. 

The theory is semiclassical. It is based on the real-time instanton approach that describes diffusion (quantum walk) over quasienergy states away from the minima of the Floquet Hamiltonian. We explicitly find the instanton trajectory. 

At the heart of the analysis are the features of the classical Hamiltonian dynamics in the rotating frame. They include the same frequencies of  the real-time  intrawell trajectories in different wells for the same quasienergy. Overall, the trajectories are double-periodic, with the complex period related to complex interwell trajectories.  

We show that the phase portrait of the trajectories in complex time displays an hourglass.  This topological feature is related to the merging of the branching points in the complex phase space. 

We express the distribution over the intrawell Floquet states and the activation energies of quantum switching in terms of the periods and the positions of the poles of the trajectories, which are all given by simple integrals. An interesting physical feature is the localization in a shallow well for certain interrelation between the parameters. Such localization occurs even though, generally,  the rate of switching from the shallow well is higher than from the deep one. This quantum effect is a consequence of the topology of the classical phase portrait. Our approach allows us to describe the effect analytically.

An important consequence of the bias-induced change of the phase-flip rates is the possibility of nontrivial many-body effects  in systems of coupled quantum parametric oscillators. If the oscillators are modulated at the same frequency $\omega_p$, the major effect of a comparatively weak coupling is the force that the oscillators exert on each other. Since the oscillators vibrate at $\omega_p/2$, this force is essentially the same as the dynamical bias we studied in this paper. Its sign, for a given oscillator, depends on the vibrational states, i.e., the phases, of the oscillators coupled to it. Consequently, the change of the phase flip rates of a given oscillator depends on the phases of the oscillators coupled to it \cite{Dykman2018,Han2024,Alvarez2024}. This maps the quantum system, in the presence of dissipation, on the Ising system. If the oscillators are nonequivalent, the system becomes nonreciprocal, as demonstrated in the classical domain for two oscillators \cite{Han2024}. The present paper shows that new quantum effects should emerge in a system of coupled oscillators. The results are thus a step toward the study of the rich dynamics of large systems of quantum oscillators, including nonequilibrium  phase transitions and  the effects of nonreciprocity in the quantum domain.


\begin{acknowledgments}
We are grateful to Rodrigo Corti\~nas for helpful comments on the paper. 
	D.\,K.\,J.\,B. and W.\,B. gratefully acknowledge financial support from the Deutsche Forschungsgemeinschaft(DFG, German Research Foundation) through Project-ID 425217212 - SFB 1432.	M.I.D. acknowledges partial support  from the  Moore Foundation (Grant No. 12214).
\end{acknowledgments}

\appendix

\section{Classical trajectories} 
\label{sec:append_trajectories}

In the WKB approximation, the rates of transitions between the intrawell states of $g(Q,P)$ can be calculated using the classical Hamiltonian equations of motion of the system with the coordinate $Q$ and momentum $P$ and the Hamiltonian $g(Q,P)$, see 
Eq.~(\ref{eq:eom}). 
We now show that the solution of these equations are given by elliptic functions. To this end, we consider the equation for $Q$ as a function of time for a given $g$,
%
\begin{align}
\label{eq:turning_points_app}
	\dot{Q} &= P(Q|g) \sqrt{B(Q)}\,,
\end{align}
where $P(Q|g)$ and $B(Q)$ are given in Eq.~(\ref{eq:turning_points}). We first change from $Q$ to 
$\tilde Q = Q - (Q_+ + Q_-)/2$, where $Q_\pm$ are the roots of the quadratic equation $B(Q)=0$. Next we change to $\tilde Q = \frac{1}{2}(Q_+ - Q_-)\cosh \xi$. It is convenient to write Eq.~(\ref{eq:turning_points_app}) for the function $x=\exp(\xi)$,
\begin{align}
\label{eq:elliptic_app}
&\dot x = 2xP\bigl(Q(x)|g\bigr),\nonumber\\
&Q(x) = \frac{1}{2}(Q_+ + Q_-) + \frac{Q_+ - Q_-}{4}\left(x+ \frac{1}{x}\right)\,,
\end{align}
[we note for completeness that $\sqrt{B(Q)} = (Q_+ - Q_-)(x-x^{-1})/2$]. 

It is seen from Eq.~(\ref{eq:turning_points}) that, with the account taken of the form of $\sqrt{B(Q)}$, the function $xP\bigl(Q(x)|g\bigr)$ is a square root of a quartic polynomial of $x$. Therefore $x(t)$ is expressed in terms of the Jacobi elliptic functions \cite{Gradstein2007}. These functions are  double-periodic, with a real and complex periods. Respectively, the functions $Q(t)$ and $P(t)$, which are rational functions of $x(t)$, are also double-periodic, and moreover, the only singularities they have in the parallelogram of periods are poles. 


\section{The interrelation between the complex period and the positions of the poles}
\label{sec:append_relation}

In this section we prove the interrelation Im~$(\tau_p^{(2)} -\tau_\ast - \tau_{\ast\ast})>0$ between the complex period $\tau_p\2$ and the positions $\tau_*, \tau_{**}$ of the poles of $Q(t)$ and $P(t)$. This interrelation shows that the function $R'(g)$ given by Eq.~(\ref{eq:R_prime}) is positive. In turn, the condition $R'>0$  means that the distribution over the intrawell states falls off monotonically with the increasing $g$. 

\begin{figure}[t]
	\includegraphics[width=1\linewidth]{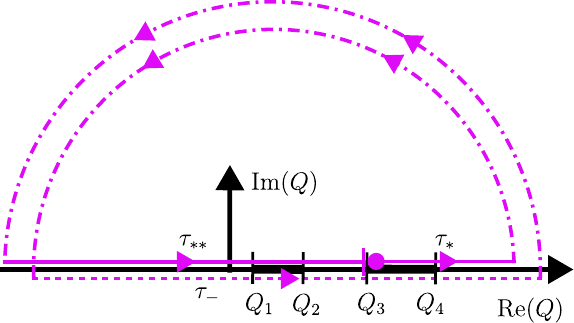}
	\caption{A periodic Hamiltonian trajectory that starts at $Q_3$ and goes twice along a semicircle $|Q|\to \infty$ in the complex-$Q$ plane.  The points $Q_i$ with $i=1,... 4$ are the turning points of the intrawell trajectories with given quasienergy $g$. }
	\label{fig:path_imaginary_period}
\end{figure}

We consider the duration of motion over the contour shown in Fig.~\ref{fig:path_imaginary_period} with a given quasienergy $g$.    
We start at the smaller-$Q$ turning point $Q_3$  of the $\sigma=1$-well  and go to $Q\to \infty$.  On this section of the contour, as explained in the main text, cf. Eq.~(\ref{eq:turning_points}),
\begin{align}
	\label{eq:append_dt}
	dt = dQ/\dot Q \equiv dQ/P(Q|g)B^{1/2}(Q)\, .
\end{align}
We take the branch $P(Q|g)$ with $P(Q|g)>0$ inside the $\sigma=1$-well, where $Q_3 < Q < Q_4$, then go around the turning point $Q_4$ in the upper half-plane of the complex-$Q$ plane, so that Im~$P<0$ for $Q>Q_4$. Then the duration of motion from $Q_3$ to $Q\to \infty$ is $\tau_\ast$.  

We then go to $Q\to -\infty$ along a semicircle  in the complex-$Q$ plane. Since $P(Q|g) \sqrt{B(Q)}\propto Q^2$, the time it takes to do so goes to zero for $|Q|\to \infty$. However,  $\sqrt{B(Q)}$ changes sign. Respectively, the momentum $P$ switches from the branch described by Eq.~(\ref{eq:turning_points}) to the branch 
\[P_-(Q|g) = [- Q^2 -1  +\mu -  B^{1/2}(Q)]^{1/2}.\]
On this branch the momentum remains complex for any real $Q$. Note that Im $P_-(Q|g)>0$,  as the sign of the momentum has changed as a result of going around the semicircle in the $Q$-plane. 

We now consider moving from $Q\to -\infty$ to $Q\to \infty$ along the real $Q$ axis. This time is given by integrating Eq.~(\ref{eq:append_dt}) with $P(Q|g)$ replaced with $P_-(Q|g)$,
\begin{align}
\label{eq:tau_-}
	\tau_- &= \int_{-\infty}^{\infty} dQ/P_-(Q|g)B^{1/2}(Q)\,,
\end{align}
Since Im~$P_-(Q|g)>0$ and $B^{1/2}<0$, the imaginary part of the integrand in Eq.~(\ref{eq:tau_-}) is positive, Im~$\tau_- > 0$.  
For $g<g_c$ in calculating $\tau_-$ we have to go around the poles of $B(Q)$ on the real-$Q$ axis at $Q=Q_\pm$, which gives a real contribution $\tau_p^{(1)}/2$ to $\tau_-$. 

From  $Q=+\infty$ we again go around a semicircle in the complex $Q$-plane. Now on the real-$Q$ axis at   $Q \to -\infty$ we have $\sqrt{B(Q)}>0$ while the momentum is given by Eq.~(\ref{eq:turning_points}) for  $P(Q|g)$ with Im~$P(Q|g)<0$. We integrate Eq.~(\ref{eq:append_dt}) along the real-$Q$ axis from $Q\to -\infty$ to $Q_3$.  It takes time $\tau_{**}(g)$, cf. Eq.~(\ref{eq:tau**}) for  $\tau_{**}(g)$.

Overall, since the trajectory comes to the initial state $Q_3$, the duration of motion is given by $\tau_p\2$. Therefore we have 
\begin{align*}
	\Im~\tau_p^{(2)} = \Im~(\tau_\ast + \tau_- + \tau_{\ast\ast})\,,
\end{align*}
and thus
\begin{align*}
	\Im~(\tau_p^{(2)} -\tau_\ast - \tau_{\ast\ast})>0\,.
\end{align*}
%

%
%
%
%


\section{Single well regime}
\label{sec:append_single_well}
\begin{figure*}[t]
	\includegraphics[width=1\linewidth]{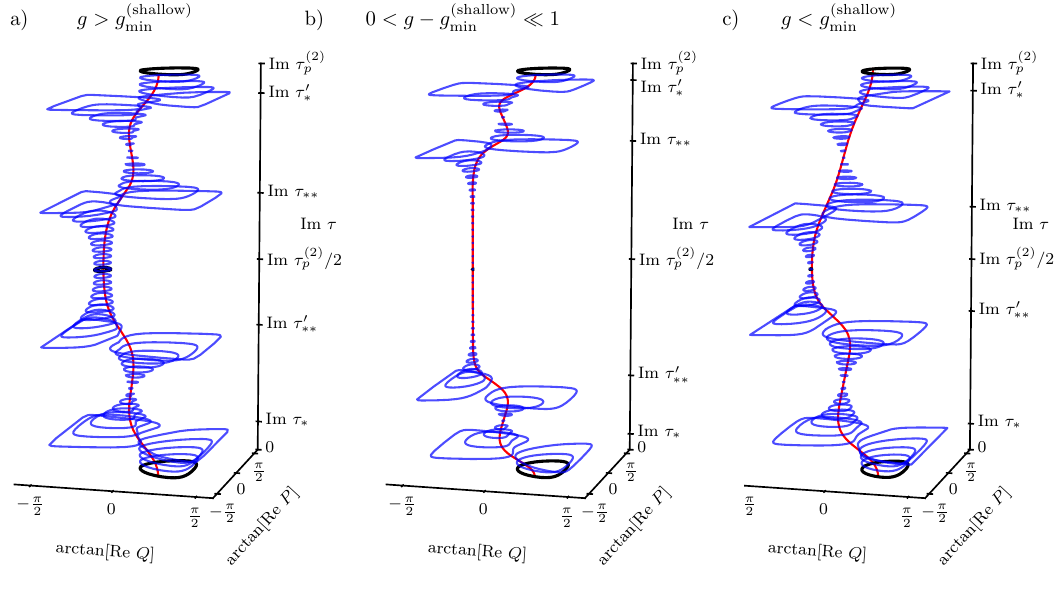}
	\caption{Phase portrait of the Hamiltonian system in the single-well regime. Panels (a) - (c) show the real-time Hamiltonian trajectories $Q(t+\tau)$, $P(t+\tau)$ (blue lines) calculated from Eq.~(\ref{eq:eom}) for different imaginary $\tau$. The trajectories connecting the turning points $Q_3$ and $Q_4$ at $\Im~\tau=0$ and $Q_1$ and $Q_2$ at $\Im~\tau=\tau_p^{(2)}/2$ are shown in black. The red lines show the  trajectories in purely imaginary time. For all plots $\mu=0.2$ and $\alpha_d=0.3$. The panels (a), (b), and (c) refer to $g=-0.01$, $g=-0.05297$, and $g=-0.1$, respectively.}
	\label{fig:topology_single_well}
\end{figure*}

Because the wells of $g(Q,P)$ have different depths, there is a region of $g$ where the vibrations occur only in the deeper well. For concreteness, Fig ~\ref{fig:sketch1}~(c) shows positions of the turning points in this case for $\alpha_d>0$, where the right,  i.e., the $\sigma=1$, well of $g(Q,P)$ is deeper than the left, i.e., the $\sigma=-1$ well. Here, as $g$ decreases, the turning points $Q_1$ and $Q_2$ of the left-well trajectories approach each other and ultimately merge at $g=g_\mathrm{min}^\mathrm{shallow}$. The minimum of the well is located at $Q_\mathrm{min}^\mathrm{shallow}$ given by the equations $P(Q|g)=0, \partial_Q g(Q,0) =0$. For $g<g_\mathrm{min}^\mathrm{shallow}$ the turning points $Q_{1,2}$ become complex. They are connected by a real time trajectory. The period of motion along this trajectory for a given $g$ coincides with the period of motion along the real-$Q$ axis in the right well for the same $g$, see Fig.~\ref{fig:sketch1}.

The complex turning points $Q_1$ and $Q_2$ and connected to the turning point $Q_3$ of the $\sigma=1$ well with the same $g$ by complex-time Hamiltonian trajectories. The time it takes is $\tau_p\2/2$, since moving back and forth is the period $\tau_p\2$, cf. Eq.~(\ref{eq:tau_p_2}). We first discuss this motion for the case where $Q_\pm$ are complex. Starting from $Q_3$ and moving in imaginary time along the real-$Q$ axis, one can reach the real-$Q$ point on the real time trajectory connecting $Q_1$ and $Q_2$, see Fig.~\ref{fig:sketch1}~(c). From there $Q_1$ or $Q_2$ can be reached by moving in real time. By symmetry, it takes the same time to reach $Q_1$ or $Q_2$, and this time is $\tau_p\1/4$, since the real time to move from $Q_1$ to $Q_2$  is $\tau_p\1/2$. Therefore, the complex period $\tau_p^{(2)}$ acquires a real part of $\tau_p^{(1)}/2$. We note that it is legitimate to appropriately deform the integration contour of $\int_{Q_3}^{Q_1} dQ/\dot Q$, which gives $\tau_p\2/2$, as long as it does not go around the branching points $Q_+$ or $Q_-$.

In the case where $Q_\pm$ are on the real axis, moving between them leads to a real contribution $\tau_p\1/2$ to  $\int_{Q_3}^{Q_1} dQ/\dot Q$ and thus to $\tau_p\2/2$. Since $Q$ and $P$ are periodic in the real time, we can always add or subtract $\tau_p^{(1)}$ from the complex period. 

In Fig. \ref{fig:topology_single_well} we show the change of the topology of the Hamiltonian trajectories as $g$ changes from the double-well to the single-well range. In contrast to Fig.~\ref{fig:topology}, shown is the evolution of the phase portrait over the whole imaginary part of the complex period  $\tau_p\2$.  As $g$ approaches the minimum of the shallow well, the imaginary period increases until it diverges for $g=g_\mathrm{min}^\mathrm{shallow}$. 
The $\sigma=1$-well trajectories become disconnected from the trajectories that connect the turning points $Q_1$ and $Q_2$.  For $g< g_\mathrm{min}^\mathrm{shallow}$ they reconnect, except that $Q_1$ and $Q_2$ are now complex. Still $Q_1$ and $Q_2$ are connected by a real-time trajectory with period $\tau_p\1$. The  period $\tau_p\2$ becomes finite for $g<g_\mathrm{min}^\mathrm{shallow}$, however it acquires a real part equal to half the real period $\tau_p^{(1)}$. Over the time Im~$\tau_p\2$ the imaginary-time trajectory shown in red returns to a point on the trajectory of the $\sigma=1$-well shifted by $\tau_p^{(1)}/2$, i.e. the sign of $P$  changes.

\section{The behavior at the minima of $g(Q,P)$}
\label{sec:append_minima}
As $g$ approaches a minimum $g_\mathrm{min}$ of one of the wells,  the corresponding turning points approach each other and for $g=g_\mathrm{min}$ merge together at $Q=Q_\mathrm{min}$.  Since at the minimum $P(Q_\mathrm{min}|g)=0$ and $\partial_Q g(Q,0)=0$, near the minimum we have $P^2(Q|g)\approx d_1(g-g_\mathrm{min}) - d_2(Q-Q_\mathrm{min})^2 $ with $d_{1,2}>0$. Therefore,  from Eq.~(\ref{eq:tau_p_2}) 
\[\mathrm{Im}\,\tau_p\2  \propto \log (g-g_\mathrm{min})\]
whether $Q_2$ or $Q_3$ approaches $Q_\mathrm{min}$, i.e., whether $g$ is near the minimum of the $\sigma=1$- or $\sigma=-1$-well. This divergence is also shown in Fig.~(\ref{fig:rprime}). 

The pole $\mathrm{Im}~\tau_{\ast\ast}$ also goes to infinity as $g$ approaches a minimum. It follows from Eq.~(\ref{eq:tau**}) that, near the minimum of the shallow well, the leading-order term in $\mathrm{Im}~\tau_{\ast\ast}$ coincides with that in $\mathrm{Im}~\tau_p^{(2)}$, whereas Im~$\tau_*$ does not diverge. On the other hand, near the minimum of the deeper well  $\mathrm{Im}~\tau_{\ast\ast}\approx \mathrm{Im}~\tau_p^{(2)}/2$, but, as seen from Eq.~(\ref{eq:tau*}), Im~$\tau_* \approx \mathrm{Im}\,\tau_p\2/2$.  Therefore, from Eq.~(\ref{eq:R_prime}), $R'(g)$  remains finite near both minima of $g(Q,P)$. This is to be contrasted with the behavior of $R'$ near $g_c$, where it diverges,  see Fig.~\ref{fig:rprime}. 

A straightforward calculation shows that, near a minimum of $g(Q,P)$
\begin{align}
\label{eq:R_prime_min}
& R'(g) \approx 2(A_P A_Q)^{-1/2}\log\frac{\sqrt{A_P} + \sqrt{A_Q}}{|\sqrt{A_P} - \sqrt{A_Q}|},\nonumber\\
 &A_P = Q_\mathrm{min}^2 -\mu +1,\quad A_Q = 3 Q_\mathrm{min}^2 -\mu -1.
 \end{align}
For $\alpha_d=0$ we have $Q_\mathrm{min} = \pm (\mu +1)^{1/2}$, and the above expression coincides with the result \cite{Marthaler2006}.For  $\alpha_d=0$ and $\mu=0$, the critical value $g_c$ as given by Eq.~(\ref{eq:critical_g}) coincides with $g_\mathrm{min} = -1/4$. In this case $R'(g)$ logarithmically diverges as $g\to g_\mathrm{min}$. This follows from Eqs.~(\ref{eq:tau_p_2}), (\ref{eq:tau*}), (\ref{eq:tau**}), and (\ref{eq:R_prime}).

We note that,  for $\alpha_d=|\mu|$, we find $g_c = g_\mathrm{min}^\mathrm{deep}$ for $-0.25<\mu<0$ and $g_c=g_\mathrm{min}^\mathrm{shallow}$ for $0<\mu<2$. For $\mu>2$ the condition that the stationary points of $g(Q,P)$ are minima with respect to $P$ is violated for the point $(Q_s,P=0)$, which is the saddle point for $\mu<2$.


%

\end{document}